\documentclass[aps,preprint,groupedaddress,nofootinbib]{revtex4-1}
\usepackage{amsmath}
\usepackage{appendix}
\usepackage{slashed}
\usepackage{amsfonts}
\usepackage{amsmath}
\usepackage{booktabs}
\usepackage{siunitx}
\usepackage{multirow}
\usepackage{colortbl}
\usepackage{xcolor}
\usepackage[%
  colorlinks=true,
  urlcolor=blue,
  linkcolor=blue,
  citecolor=blue
  ]{hyperref}

\def\GG{\langle g_s^2 G^2 \rangle}
\def\sp{\langle \bar s s \rangle}

\usepackage{graphicx}

\begin{document}

\title{ARE THE NEW EXCITED $\Omega_c$ BARYONS NEGATIVE PARITY STATES?}

\author{T.~M.~Aliev}
\email[]{taliev@metu.edu.tr}
\author{S.~Bilmis}
\email[]{sbilmis@metu.edu.tr}
\author{ M.~Savci}
\email[]{savci@metu.edu.tr}
\affiliation{Department of Physics, Middle East Technical University, 06800, Ankara, Turkey}

\begin{abstract}
We calculate the mass and residue of the newly observed $\Omega_c(3000)$ and $\Omega_c(3066)$ states with quantum numbers $J^P = \frac{1}{2}^{-}$ and $\frac{3}{2}^{-}$ within QCD sum rules. The calculation is carried out by using the general form for interpolating curent for $J = \frac{1}{2}$ baryon. Our predictions on masses are in good agreement with the experimental results.

\end{abstract}


\maketitle

\section{Introduction}
\label{sec:intro}
LHCb Collaboration observed five very narrow excited $\Omega_c$ baryons decaying into $\Xi^+ \bar{K}$~\cite{Aaij:2017nav}. The masses and decay widths of these new states are:
\begin{equation}
  \label{eq:1}
  \begin{split}
    \Gamma_1 &= 4.5 \pm 0.6 \pm 0.3~(\rm{MeV}),~~~m_1 = 3000.4 \pm 0.2 \pm 0.1~{\rm{MeV}}, \\
    \Gamma_2 &= 0.8 \pm 0.2 \pm 0.1~(\rm{MeV}),~~~m_2 = 3050.2 \pm 0.1 \pm 0.1~{\rm{MeV}}, \\
    \Gamma_3 &= 3.5 \pm 0.4 \pm 0.2~(\rm{MeV}),~~~m_3 = 3065.6 \pm 0.1 \pm 0.3~{\rm{MeV}}, \\
    \Gamma_4 &= 8.7 \pm 1.0 \pm 0.8~(\rm{MeV}),~~~m_4 = 3090.2 \pm 0.3 \pm 0.5~{\rm{MeV}}, \\
    \Gamma_5 &= 1.1 \pm 0.8 \pm 0.4~(\rm{MeV}),~~~m_5 = 31191.1 \pm 0.3 \pm 0.9~{\rm{MeV}}. \\
  \end{split}
\end{equation}
These states except $\Omega_c(3119)$ have also been confirmed by BELLE Collaboration~\cite{Yelton:2017qxg} and masses as well as the relative branching ratios of the hadronic decays of them are measured~\cite{Yelton:2017uzv}. 
However, quantum numbers ($J^P$) of these new states have not been established in the experiments yet. Hence, in recent studies, various scenarios have been employed concerning the nature of these states. The spectra of the newly discovered $\Omega_c$ baryons within different approaches such as QCD sum rules \cite{Wang:2017zjw,Wang:2017xam,Agaev:2017jyt},
chiral perturbation theory \cite{PhysRevD.95.094018},
chiral quark soliton model \cite{Kim:2017jpx}, and
heavy quark + light diquark framework \cite{Wang:2017vnc} have been widely discussed in the literature.
For instance, in Ref.~\cite{Agaev:2017jyt}, the two states with masses $m_3$ and $m_5$ are assumed to have the $J^P = \frac{1}{2}^+$ and $\frac{3}{2}^+$ quantum numbers which are radial excitations of ground state $\Omega_c$ and $\Omega_c^*$ baryons, and within QCD sum rules their masses are estimated. In~\cite{Karliner:2017kfm}, the authors try to answer the following questions: Why are the five states discovered? Why are they narrow? What are their spin-parity quantum numbers? Do similar states of other heavy baryons, as well as $\Omega_c$, exist for beauty baryons within the quark model? The authors of~\cite{Karliner:2017kfm} assumed these states as bound states of a c-quark and a \rm{P} wave \rm{ss}-diquark. This picture predicts the existence of five states with negative parity.\footnote{In this work, the authors also present an alternative possibility that the two heavy states are $2S$ excitations with $J^P = \frac{1}{2}^{+}$ and $J^P = \frac{3}{2}^{+}$, while the three light states are interpreted as $J^P = \frac{3}{2}^{-}$, $\frac{3}{2}^{-}$ and $\frac{5}{2}^{-}$ states. } Additionally, the ground and excited state spectra of $\Omega_c^0$ baryons are analyzed from lattice QCD in~\cite{Padmanath:2017lng} and the result strongly indicated that the states $\Omega_c^0(3000)$, $\Omega_c^0(3050)$ and $\Omega_c^0(3066)$, $\Omega_c^0(3090)$ and  $\Omega_c^0(3119)$ have parity spin $J^P = \frac{1}{2}^{-}$, $\frac{3}{2}^{-}$ and $\frac{5}{2}^{-}$ respectively.

Moreover, the strong and radiative decays of $\Omega_c$ baryons are very promising to establish the quantum numbers of these states. In this regard, several decay modes of these hadrons are analyzed within different methods such as the constituent quark model \cite{Wang:2017hej}, quark pair creation model ($^3P_0$-model \cite{PhysRevD.95.114024,Chen:2017gnu}), chiral quark model \cite{Huang:2017dwn,Wang:2017vnc}, and light-cone QCD sum rules \cite{Aliev:2018uby,PhysRevD.95.094008}. Newly observed $\Omega_c$ baryon as pentaquarks within the chiral quark model is discussed in~\cite{Yang:2017rpg} and it is shown that the $\Xi^- \bar{D},~\Xi_c \bar{K}$, and $\Xi_c^* \bar{K}$ are the possible decay candidates of these new particles. 


Following the work \cite{Karliner:2017kfm}, we assume that the newly observed $\Omega_c$ baryons are negative parity baryons and in present letter within QCD sum rules method, we estimate the mass and residues of the $J^P=\frac{1}{2}^-$, $J^P=\frac{3}{2}^-$ states respectively. The paper is organized as follows. In section~\ref{sec:2}, we derive the mass sum rules for negative parity $\Omega_c^0(3000)$ and $\Omega_c^0(3066)$ with $J^P=\frac{1}{2}^-$ and $J^P=\frac{3}{2}^-$. Section~\ref{sec:numeric} is devoted to the numerical analysis of the obtained sum rules. The last section contains discussions and conclusion.

\section{ Mass sum rules for $\Omega_c(3000)$ and $\Omega_c(3066)$ baryons}
\label{sec:2}
To derive the sum rules for the mass and residues of the $\Omega_c(3000)$ and $\Omega_c(3060)$ states we consider the following two-point correlation functions
\begin{equation}
  \label{eq:2}
  \Pi_{(\mu\nu)}(p) = \int d^4 x e^{i p x} \big\{ \langle 0| T\{\eta_{Q_{(\mu)}}(x) \bar{\eta}_{Q_{(\nu)}}(0) \} |0 \rangle \big\}
\end{equation}
where
\begin{equation}
  \label{eq:3}
  \begin{split}
    \eta_{Q} = \frac{1}{\sqrt{2}} \epsilon^{abc} \bigg\{& (s^{a^T} C Q^b) \gamma_5 C^c - (Q^{a^T} C s^b) \gamma_5 s^c \\
    & + \beta \big[ (s^{a^T} C \gamma_5 Q^b) s^c - (Q^{a^T} C \gamma_5 s^b) s^c \big] \bigg\}
  \end{split}
\end{equation}
and
\begin{equation}
  \label{eq:3}
  \begin{split}
    \eta_{Q_\mu} = \frac{1}{\sqrt{3}} \epsilon^{abc} \bigg\{& (s^{a} C \gamma_\mu s^b) Q^c + (s^{a} C \gamma_\mu Q^b) s^c  +  (Q^{a} C \gamma_\mu s^b) s^c ) \bigg\}
  \end{split}
\end{equation}
are the interpolating currents of $\Omega_Q$ baryons with $J^P = \frac{1}{2}^+$ and $\frac{3}{2}^+$ (see for example~\cite{Bagan:1991sc}). In the expressions of the currents, \rm{a,b,c} are the color indices, \rm{C} is the charge conjugation operator, \rm{Q} is the heavy \rm{c} quark and $\beta$ is the arbitrary parameter, where $\beta=-1$ corresponds to so-called Ioffe current.

In order to obtain the mass sum rules, the correlation functions are calculated in terms of hadrons and quark-gluon degrees of freedom. Then with the help of dispersion relation, these results are equated. In this way, the mass sum rules are obtained.

It should be noted that the interpolating currents $\eta_Q$ and $\eta_{Q_\mu}$interact with both positive and negative parity baryons. Using this fact and saturating eq.~(\ref{eq:2}) with positive and negative parity baryons we get
\begin{equation}
  \label{eq:4}
  \begin{split}
    \Pi(p) &= \frac{\langle 0 |\eta_Q | \Omega_Q^{(+)} \rangle \langle \Omega_Q^+ | \bar{\eta}_Q |0 \rangle}{m_{\Omega^+}^2 - p^2} \\
    &+ \frac{\langle 0 |\eta_Q | \Omega_Q^{(-)} \rangle \langle \Omega_Q^- | \bar{\eta}_Q |0 \rangle}{m_{\Omega^{-}}^2 - p^2} + ... \\
      \Pi_{\mu \nu}(p) &= \frac{\langle 0 |\eta_{Q_\mu} | \Omega_Q^{*(+)} \rangle \langle \Omega_Q^{*+} | \bar{\eta}_\nu |0 \rangle}{m_{\Omega_Q^{+*}}^2 - p^2} \\
      &+ \frac{\langle 0 |\eta_{Q_\mu} | \Omega_Q^{*(-)} \rangle \langle \Omega_Q^{*-} | \bar{\eta}_\nu |0 \rangle}{m_{\Omega^{*-}}^2 - p^2} + ...  
  \end{split}
\end{equation}

Here, $\Omega_Q^+~(\Omega^{+*})$, $\Omega_Q^{-}~(\Omega^{-*})$ are the ground states positive and negative parity baryons with spin-$1/2~(3/2)$, respectively. Moreover, for briefness, we will denote the mass of the negative parity spin $\frac{1}{2}(\frac{3}{2})$ $\Omega_Q$ baryons as $m_{-} (m_{-}^*)$. The dots describe for higher states and continuum contributions.


The matrix elements entering to eqs.~(\ref{eq:3}) and (\ref{eq:4}) are determined as follows:
\begin{equation}
  \label{eq:6}
  \begin{split}
    \langle 0 | \eta_Q | (\frac{1}{2}^+) \rangle &= \lambda_+ u(p) \\
    \langle 0 | \eta_Q | (\frac{1}{2}^-) \rangle &= \lambda_{-} \gamma_5 u(p) \\
    \langle \eta_{Q\mu} | 3/2^+ \rangle &= \lambda_{+}^{*} u_\mu(p) \\
    \langle \eta_{Q\mu} | 3/2^- \rangle &= \lambda_{-}^{*} \gamma_5 u_\mu(p)     
  \end{split}
\end{equation}
where $u_\mu(p)$ is the Rarita-Schwinger spinor for spin $3/2$ particle.


Using these matrix elements and performing summation over the spins of baryons, we get the phenomenological part of the correlation functions as
\begin{equation}
  \label{eq:8}
  \begin{split}
    \Pi(p) &= \frac{\lambda_{+}^2 (\slashed{p} + m_{+})}{m_{+}^2 - p^2} + \frac{\lambda_{-}^2 (\slashed{p} - m_{-})}{m_{-}^2 - p^2} + ...\\
    \Pi_{\mu \nu}(p) &= \bigg[ \frac{\lambda_{+}^{*^2} (\slashed{p} + m_{+}^*)}{m_{+}^{*2} - p^2} + \frac{\lambda_{-}^2 (\slashed{p} - m_{-}^{*})}{m_{-}^{*2} - p^2} \bigg] g_{\mu \nu}     
  \end{split}
\end{equation}

Notice that, the states with masses $m_{\Omega_c}^+ = 2695~\rm{MeV}$, $m_{\Omega_c}^{*+} = 2766~\rm{MeV}$ 
are denoted as $m_{+}$, $m_{+}^{*}$ and their residues are denoted as $\lambda_{+}$ and $\lambda_{+}^{*}$ correspondingly. In these expressions, the second terms in RHS of eq.~(\ref{eq:8}) describe contributions of $\Omega(3000)$ and $\Omega_c(3066)$ states.

Here we would like to make the following remark. First of all, the summation over spin for Rarita-Schwinger spinors is performed by using formula
\begin{equation}
  \label{eq:9}
  \begin{split}
    \sum_s u_\mu(p,s) \bar{u}_\nu(p,s) = -(\slashed{p} + m) \big[ g_{\mu \nu} - \frac{1}{3}\gamma_\mu \gamma_\nu - \frac{2}{3m^2} p_\mu p_\nu +\frac{1}{3m} (p_\mu \gamma_\nu - p_\nu \gamma_\mu)\big].
  \end{split}
\end{equation}
Moreover, the interpolating current $\eta_\mu$ couples not only to the $J^P = \frac{3}{2}$ state but also to the  $\frac{1}{2}$ state. The contribution of $J^P=\frac{1}{2}$ state is determined as
\begin{equation}
  \label{eq:10}
  \begin{split}
    \langle 0 |\eta_\mu | \frac{1}{2} (p) \rangle = \big[ A p_\mu + B \gamma_\mu \big] u(p).
  \end{split}
\end{equation}
From this expression 
it follows that the structures proportional to $p_\mu$ and $\gamma_\mu$ contain contributions from $1/2$ states and it follows from eq.~(\ref{eq:9}) that only $\sim g_{\mu \nu}$ structure contains the contribution of $\frac{3}{2}$ states. For this reason in the next discussion, we choose only structures $\sim \slashed{p} g_{\mu \nu}$ or $g_{\mu\nu}$ in order to analyze the mass and residues of the spin $\frac{3}{2}$ states.

Now let us turn our attention to the calculation of the correlator function from QCD side by using the operator product expansion (OPE). For performing calculation, we need the expression of light (strange quark) and heavy quark propagators. Up to dimension eight operators, the expression of the light quark propagator in $x$ representation is given in~\cite{Aliev:2002ra,Wang:2008vg} 

\begin{equation}
  \label{eq:12}
  \begin{split}
    S_s^{ab}(x) &= \frac{i \slashed{x} \delta^{ab}}{2 \pi^2 x^4} - m_s \frac{\delta^{ab}}{{4 \pi^2 x^2}} - \frac{\delta^{ab}}{12} \langle \bar{s}s \rangle (1- \frac{i}{4} i \slashed{x}) \\
    & - \frac{x^2}{192} m_0^2 \langle \bar{s}s \rangle (1 - \frac{i}{6} \slashed{x} m_s) - \frac{x^4 \delta^{ab}}{2^9 3^3} \langle \bar{s} s \rangle \langle g_s^2 G^2 \rangle \\
    & + \frac{i}{2^5 \pi^2 x^2} (g_s G_{\alpha \beta}^n) (\slashed{x} \sigma^{\alpha \beta} + \sigma^{\alpha \beta} \slashed{x}) \frac{(\lambda^{n})^{ab}}{2} \\
    &+ \frac{1}{2^5 \pi^2} m_s \big( \ln{(-\frac{x^2 \Lambda^2}{4})} + 2\gamma_E \big) (g_s G_{\alpha \beta}^n)\frac{(\lambda^{n})^{ab}}{2} \sigma^{\alpha \beta} 
  \end{split}
\end{equation}

The vacuum expectation values of the quark and gluon field product also give contribution to the quark propogator. This matrix element is determined in following way~\cite{Wang:2008vg}.
\begin{equation}
  \label{eq:19}
  \begin{split}
    \langle 0| T\{ q_i^a \bar{q}_k^b G_{\alpha \beta}^n \} |0 \rangle &= \frac{1}{2^6 3} m_0^2 \langle \bar{q}q \rangle (\sigma_{\alpha \beta})_{ik} (\frac{\lambda^n}{2})^{ab}  \\
    &-\frac{i}{2^8 3} m_q m_0^2 \langle \bar{q} q \rangle \big[ \slashed{x} \sigma_{\alpha \beta} + \sigma_{\alpha \beta} \slashed{x} \big]_{ik} (\frac{\lambda^n}{2})^{ab} \\
    &+ \frac{x^2}{2^{10} 3^2} \langle g_s^2 G^2 \rangle \langle \bar{q} q \rangle (\sigma_{\alpha \beta})_{ik} (\frac{\lambda^n}{2})^{ab}
   \end{split}
\end{equation}

For the heavy quark propogator we employ following expression~\cite{Aliev:2008sk}
\begin{equation}
  \label{eq:13}
  \begin{split}
    S_Q^{ab}(x) &= \frac{m_Q^3 \delta^{ab}}{2\pi^2} \bigg\{ \frac{m_Q i \slashed{x}}{m_Q^2(\sqrt{-x^2})^2} K_2 (m_Q \sqrt{-x^2}) + \frac{1}{m_Q \sqrt{-x^2}} K_1 (m_Q \sqrt{-x^2}) \bigg\} \\
    &- \frac{m_Q g_s G_{\mu \nu}^{ab}}{8 (2 \pi)^2} \bigg\{ i(\sigma_{\mu \nu} \slashed{x} + \slashed{x} \sigma_{\mu \nu} ) \frac{1}{m_Q \sqrt{-x^2}} K_1 (m_Q \sqrt{-x^2}) 
    + 2\sigma_{\mu \nu} K_0 (m_Q \sqrt{-x^2}) \bigg\} \\
    &- \frac{\delta^{ab} \langle g_s^2 G^2 \rangle}{576 (2 \pi)^2 m_Q} \bigg\{ (im_Q \slashed{x} -6) m_Q \sqrt{-x^2} K_1 (m_Q \sqrt{-x^2}) + (m_Q \sqrt{-x^2})^2 K_2 (m_Q \sqrt{-x^2}) \bigg\}, 
  \end{split}
\end{equation}
where $K_n(m_Q \sqrt{-x^2})$ is the modified Bessel function of the second kind.

Using these expressions for the propagators of the heavy and light quarks, the correlation function can be calculated. Separating the coefficients of $\slashed{p}$ and $I$ operator structures in spin $1/2$ baryons and $\slashed{p} g_{\mu \nu}$, $g_{\mu \nu}$ Lorentz structures for spin $3/2$ baryons and performing Borel transformations over $p^2$ we get the following sum rules:
\begin{equation}
  \label{eq:14}
  \begin{split}
    \lambda_{+}^2 e^{- m_{+}^2/M^2} + \lambda_{-}^2 e^{- m_{-}^2/M^2} &= \Pi_1^{B} \\ 
    m_{+}\lambda_{+}^2 e^{- m_{+}^2/M^2} - m_{-}\lambda_{-}^2 e^{- m_{-}^2/M^2} &= \Pi_2^{B}  
  \end{split}
\end{equation}
and
\begin{equation}
  \label{eq:15}
  \begin{split}
    \lambda_{+}^{*^2} e^{- m_{+}^{*^2}/M^2} + \lambda_{-}^{*^2} e^{- m_{-}^{*^2}/M^2} &= \Pi_1^{*{B}} \\ 
  m_{+}^*  \lambda_{+}^{*^2} e^{- m_{+}^{*^2}/M^2} - m_{-}^* \lambda_{-}^{*^2} e^{- m_{-}^{*^2}/M^2} &= \Pi_2^{*{B}} 
  \end{split}
\end{equation}

The expressions of the invariant functions $\Pi_1^{B}$, $\Pi_2^{B}$, $\Pi_1^{*{B}}$ and $\Pi_2^{*{B}}$ are presented in Appendix~\ref{sec:appendix}.

Solving eq.(\ref{eq:14}) for the mass and residue of the spin $1/2$ state we obtain:
\begin{equation}
  \label{eq:16}
  \begin{split}
    m_{-}^2 &= \frac{d(\frac{-1}{M^2})[m_{+} \Pi_1^B - \Pi_2^B ]}{m_{+} \Pi_1^B - \Pi_2^B}, \\
    \lambda_{-}^2 &= \frac{e^{m_{-}^2/M^2} }{m_{+} + m_{-}} [m_{+} \Pi_1^B - \Pi_2^B].
  \end{split}
\end{equation}
Moreover, equations for the determination of the negative parity spin $3/2$ states formally can be obtained from eq.(\ref{eq:16}) replacing $m_{\pm} \rightarrow m_{\pm}^{*}$, $\Pi_i^B \rightarrow \Pi_i^{*B}$. We take the values of mass of ground state positive parity baryons obtained from mass sum rule, namely $m_+ = (2.685 \pm 0.123)~\rm{GeV}$, $m_+^* = (2.77 \pm 0.20)~\rm{GeV}$ and  (see for example~\cite{Aliev:2008sk,Aliev:2010yx}).
\section{Numerical Analysis}
\label{sec:numeric}
In this section, we perform numerical analysis of the mass sum rules for negative parity $\Omega_c$ and $\Omega_c^*$ baryons. The sum rules involve numerous input parameters. For numerical analysis we used the following values for the input parameters:
\begin{equation}
  \label{eq:17}
  \begin{split}
    m_c &= (1.27 \pm 0.03)~\rm{GeV}, \\
    m_s &= (96^{+8}_{-4})~\rm{MeV}, \\
    \langle \bar{s}s \rangle &= 0.8 (-0.24 \pm 0.01)^3~\rm{GeV^3}, \\
    m_0^2 &= (0.8 \pm 0.2)~\rm{GeV^2}, \\
    \langle \frac{\alpha_s}{\pi} G^2 \rangle &= (0.012 \pm 0.004)~\rm{GeV^4}
  \end{split}
\end{equation}
Note that the $\overline{\text{MS}}$-scheme is chosen for the charm quark mass which leads to reasonable suppression of $O(\alpha_s)$ radiative corrections in the perturbative part of the sum rules.
In addition to these input parameters, the sum rules contain three auxiliary parameters: Borel mass square $M^2$, continuum threshold $s_0$ and arbitrary parameter $\beta$ in the expression of the interpolating current $\eta$ for $J^P = \frac{1}{2}$ states. For the prediction of the mass of the negative parity $\Omega_c$ and $\Omega_c^*$ baryons we need to find the working region of these parameters in such a way that the mass is practically independent of them. The working region of $M^2$ is determined as follows. The lower bound of $M^2$ is obtained from the condition that the higher dimension operators contributions should be less than perturbative contribution in order to guarantee the convergence of OPE series. On the other hand, the upper bound of $M^2$ is determined from the condition that the contributions of the continuum and higher states should be less than say half of the pole contributions. Our analysis shows that these conditions are satisfied if Borel mass parameter lies in the region $2~\rm{GeV^2} \leq M^2 \leq 5~\rm{GeV^2}$ for both baryons.

\begin{figure}[hbt]
  \centering
  \includegraphics[scale=0.7]{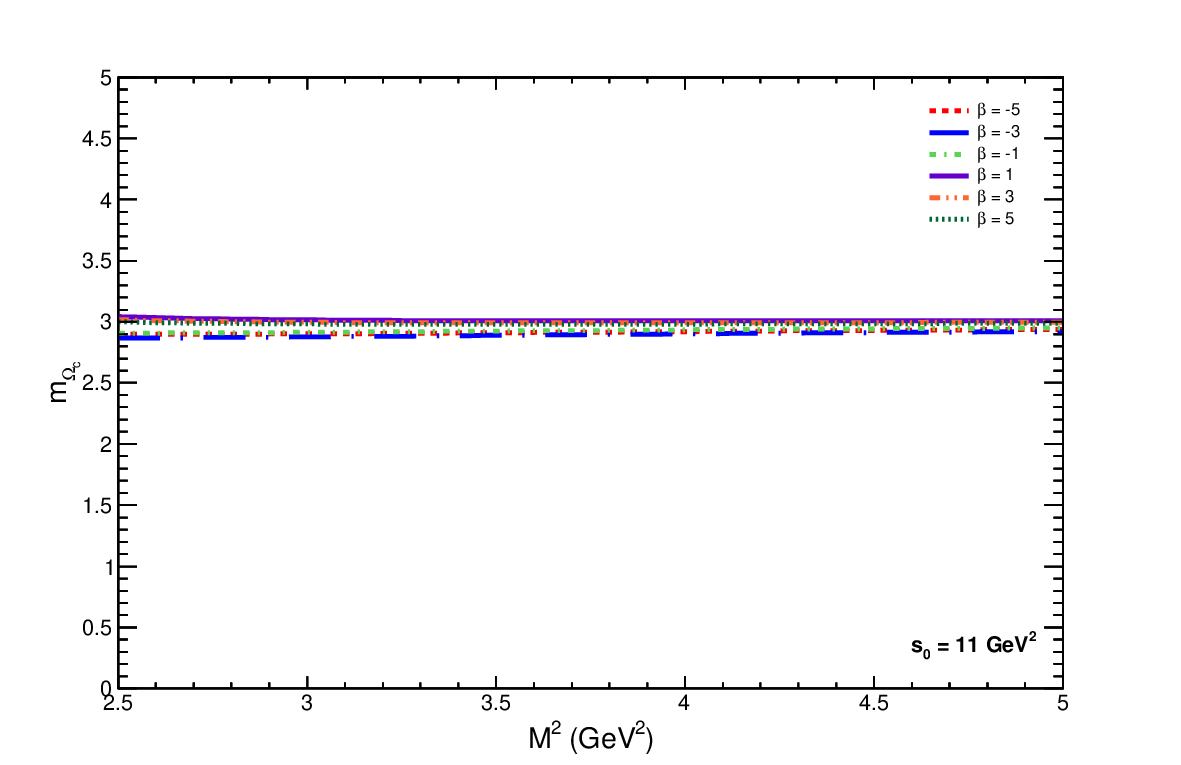}  
  \caption{The dependence of the mass of the negative parity $\Omega_c$ baryon on Borel mass parameter $M^2$ at $s_0 = 11~\rm{GeV^2}$ and for several fixed values of $\beta$ is depicted. }
  \label{fig:1}
\end{figure}

\begin{figure}[hbt]
  \centering
  \includegraphics[scale=0.7]{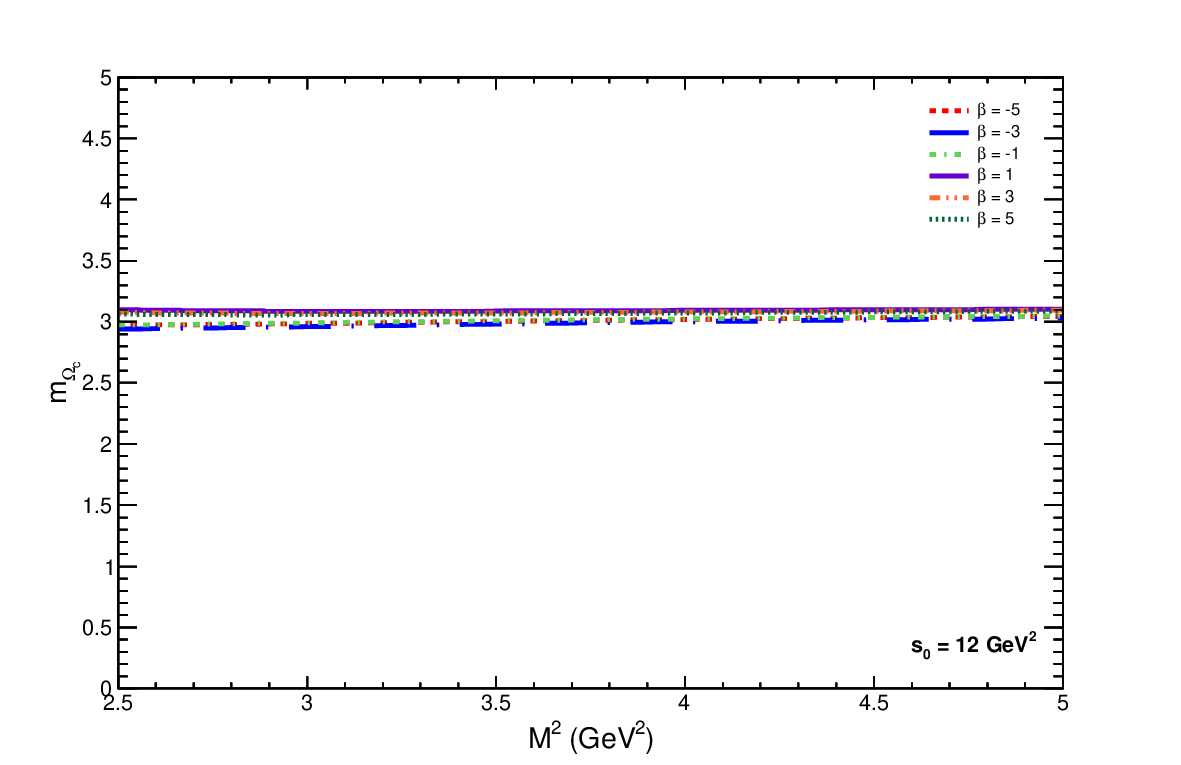}  
  \caption{Same as in Fig.~(\ref{fig:1}), but at $s_0 = 12~\rm{GeV^2}$.}
  \label{fig:2}
\end{figure}

\begin{figure}[hbt]
  \centering
  \includegraphics[scale=0.7]{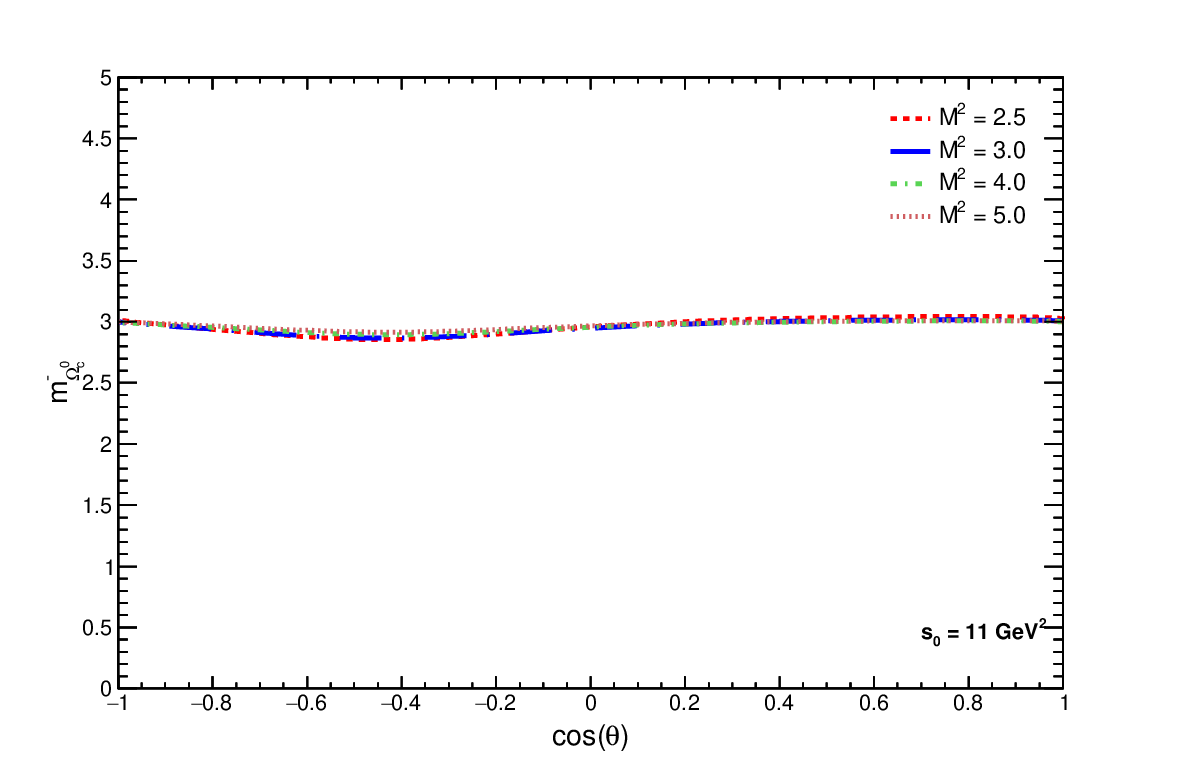}  
  \caption{The dependence of $m_{\Omega_c}$ on $\cos{\theta}$ at $s_0 = 11\rm{GeV^2}$ and at several fixed values of $M^2$.}
  \label{fig:3}
\end{figure}

The other auxiliary parameter of the sum rules is the continuum threshold $s_0$. This parameter is not totally arbitrary and is related to the energy of the first excited state. The difference $\Delta = \sqrt{s_0} - m_{ground}$ is the energy to excite the particle to its first energy state. Analysis of various sum rules shows that this parameter $\Delta$ varies between $0.3~\rm{GeV}$ and $0.8~\rm{GeV}$. In other words, $s_0$ varies in the region $(m_{\text{ground}} + 0.3~\rm{GeV})^2 \leq s_0 \leq (m_{\text{ground}} + 0.8~\rm{GeV})^2$. In performing numerical analysis, we will use values $s_0$ from this domain. Having the values of the parameters $M^2$ and $s_0$ our last attempt is to find a region of $\beta$ where the mass of $\Omega_c$ baryons become practically independent on $\beta$. In Figs.~(\ref{fig:1}) and (\ref{fig:2}), we present the dependence of $m_{-}^2$ on $M^2$ at fixed values of $s_0 = 11~\rm{GeV^2}$ and $12~\rm{GeV^2}$ at several fixed values of $\beta$, respectively. From the figure, it follows that the mass $m_{-}$ shows good stability to the variables of $M^2$ in the working region. In Fig.~(\ref{fig:3}) we present the dependence of $m_{-}$ on $\cos \theta$, where $\beta = \tan{\theta}$, at $s_0 = 11~\rm{GeV^2}$ and several fixed values of $M^2$. From this figure, we observe that $m_{-}$ becomes practically insensitive to the variation of $\cos{\theta}$ if it lies in the domain $(-1; 1)$.

\begin{figure}[hbt]
  \centering
  \includegraphics[scale=0.7]{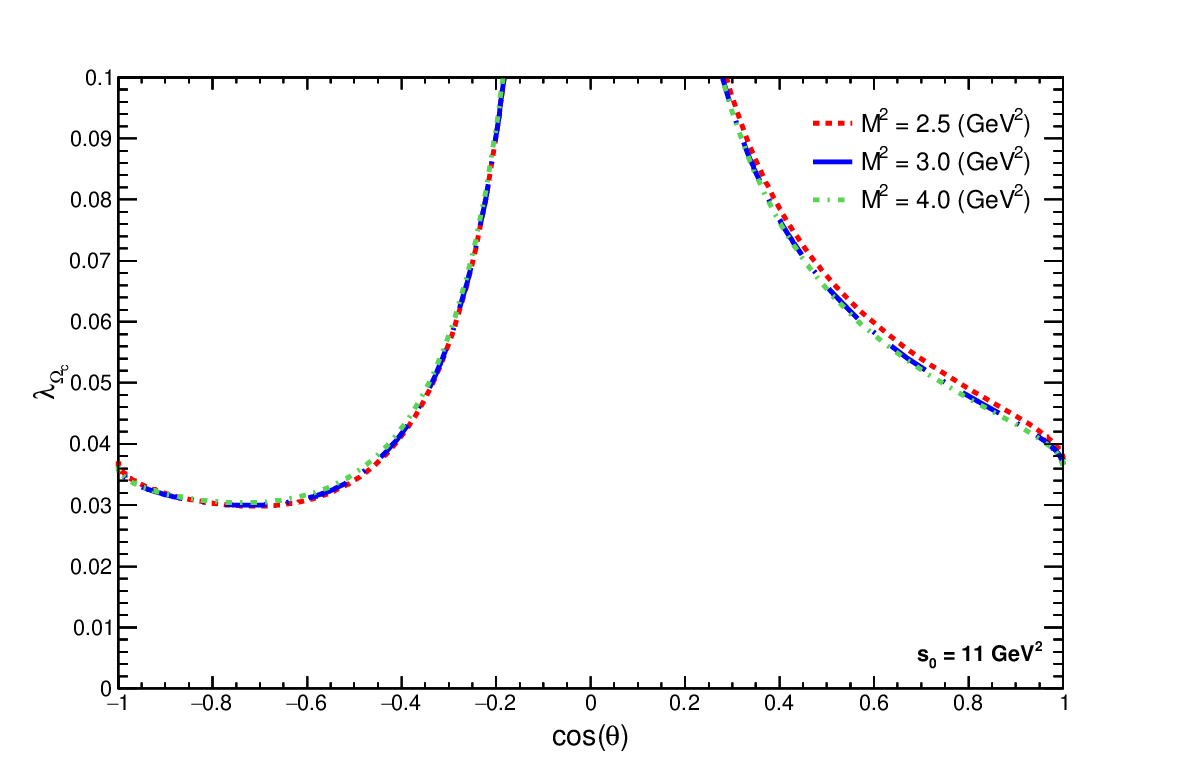}  
  \caption{The dependence of the residue ($\lambda_{\Omega_c}$) on $\cos{\theta}$ at $s_0 = 11~\rm{GeV^2}$ for three fixed values of $M^2$ is shown.}
  \label{fig:4}
\end{figure}

\begin{figure}[hbt]
  \centering
  \includegraphics[scale=0.7]{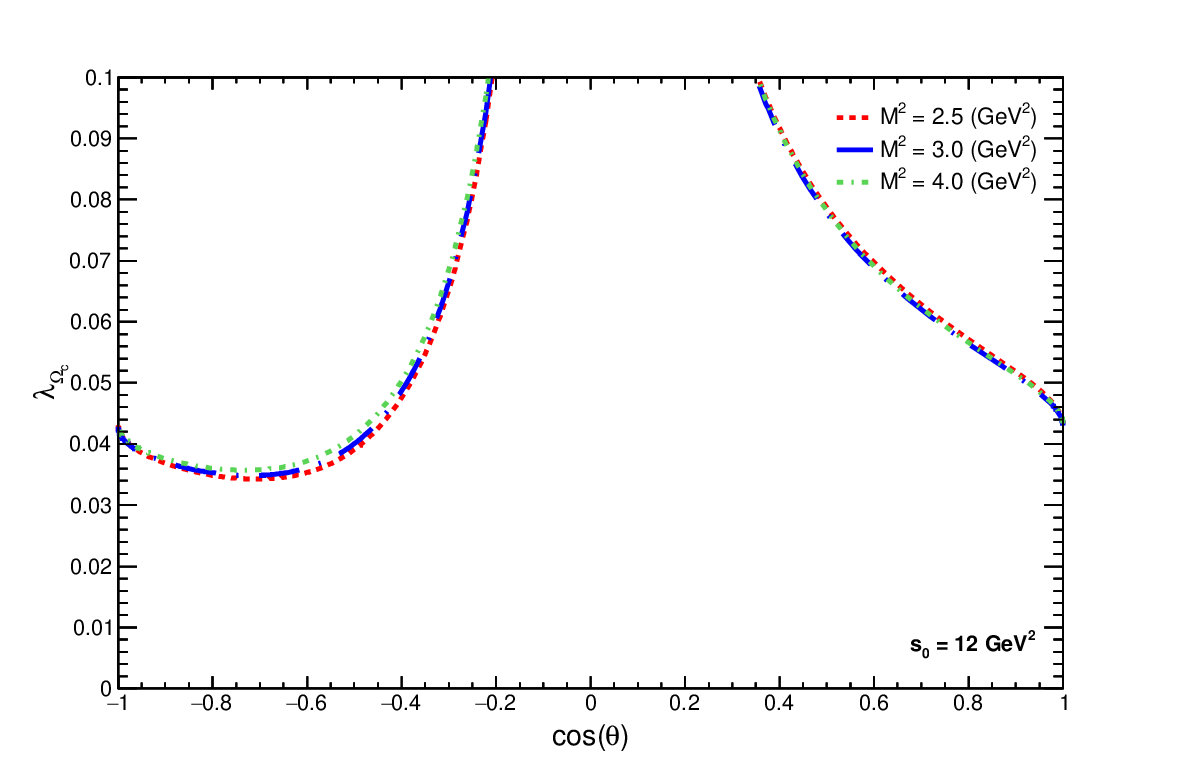}  
  \caption{Same as in Fig.~(\ref{fig:3}), but at $s_0 = 12~\rm{GeV^2}$}.
  \label{fig:5}
\end{figure}

\begin{figure}[hbt]
  \centering
  \includegraphics[scale=0.7]{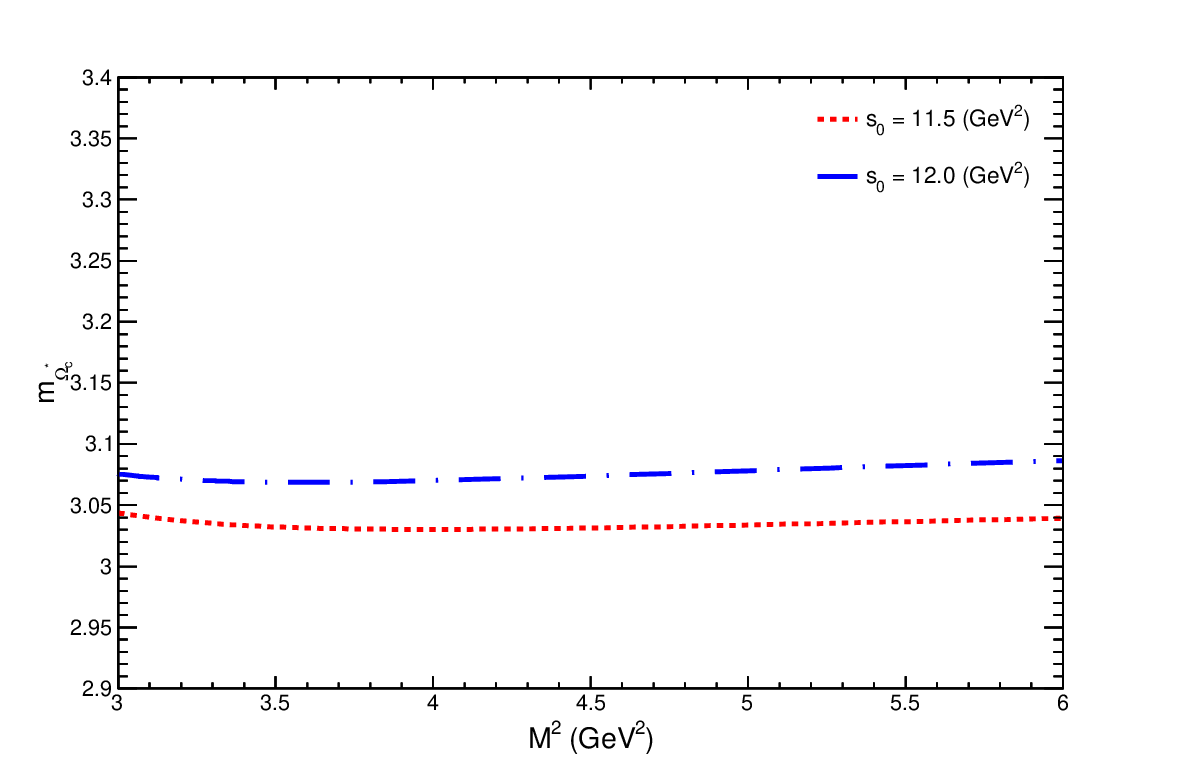}  
  \caption{The dependence of $J^P=\frac{3}{2}^-$ $\Omega_c^*$ baryon mass on $M^2$ at two fixed values of $s_0$ is depicted.}
  \label{fig:6}
\end{figure}

\begin{figure}[hbt]
  \centering
  \includegraphics[scale=0.7]{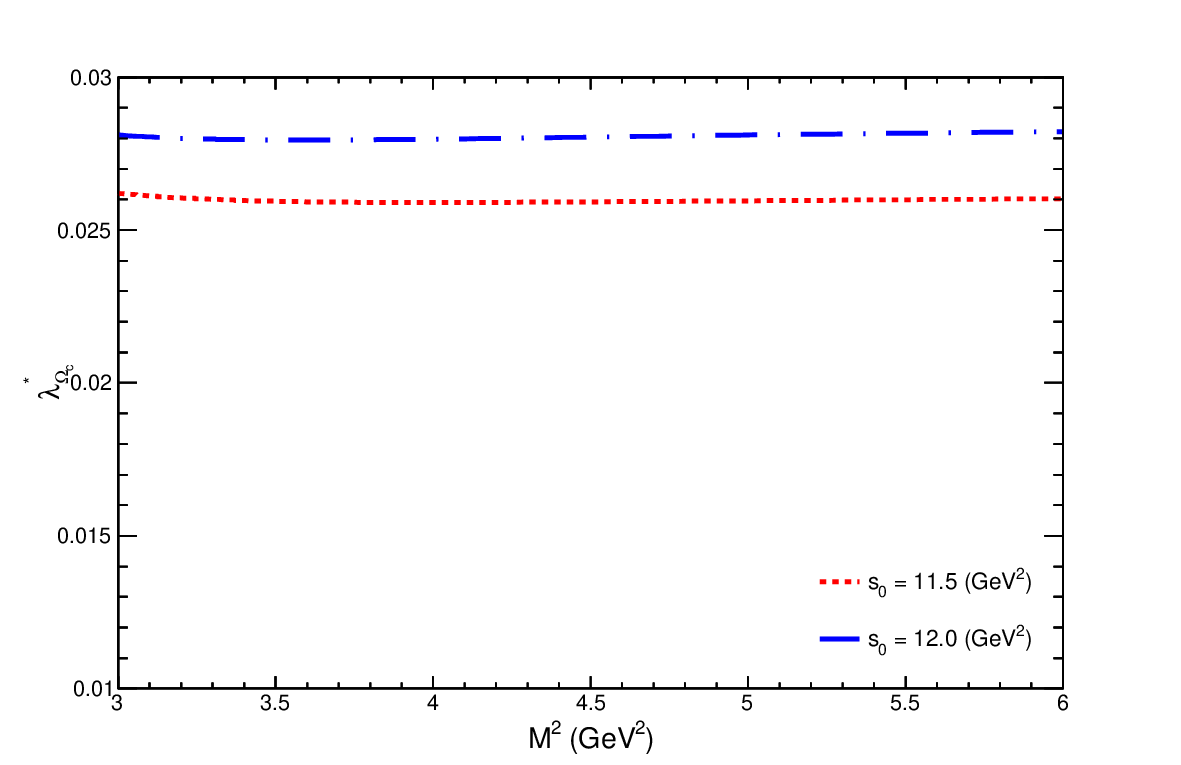}  
  \caption{Same as in Fig.~(\ref{fig:5}) but for the residue of $J^P = \frac{3}{2}^-$ $\Omega_c^{*}$ baryon.}
  \label{fig:7}
\end{figure}

In Figs~(\ref{fig:4}) and (\ref{fig:5}), we present the dependence of $\lambda_{-} $ on $\cos{\theta}$ at two fixed values of $s_0 = 11 ~\rm{GeV^2}, 12 ~\rm{GeV^2}$ and three fixed values of $M^2$ from its working region. From these Figures, we see that the $\lambda_{-}$ exhibits very good stability to the variation of $\cos{\theta}$ when it changes in the domain $ -1 < \cos{\theta} < -0.5 $ . In result, we obtained that the common working region of $\cos{\theta}$ for mass sum rules is $(-1; -0.5)$. With these findings, our final results for the mass and residue of $\Omega_c$ state are
\begin{equation}
  \label{eq:18}
  \begin{split}
    m_{-} &= (3.00 \pm 0.01)~\rm{GeV}, \\
    \lambda_{-} &= (0.036 \pm 0.007)~\rm{GeV^3}. 
  \end{split}
\end{equation}
We also perform a similar analysis for spin $3/2$ states. In Fig.~(\ref{fig:6}) we present the $M^2$ dependence of $m_{-}^*$ at two fixed values of $s_0$. We observe that in the working region of $M^2$ we have good stability of $m_{-}^*$ with respect to the variation of $M^2$. In Fig.~(\ref{fig:7}) the dependence of the residue $\lambda_{-}^{*}$ on $M^2$ at two fixed values of $s_0$ is depicted. From these figures we deduce following results: 
\begin{equation}
  \label{eq:20}
  \begin{split}
    m_{-}^{*} &= (3.06 \pm 0.02) \rm{GeV},\\ 
    \lambda_{-}^{*} &= (0.027 \pm 0.001)~\rm{GeV^3}. 
  \end{split}
\end{equation}

From comparison of our results on mass $m_{-}$ and $m_{-}^*$ are compared with the experimental data, we observed impressive agreement between them.

  Finally, we would like to note that both spectroscopic analysis and decay widths studies are crucial for the determination of the quantum states of these newly discovered baryons.
 In Ref.\cite{Aliev:2018uby}, the decay widths of $\Omega_c^0$ baryons within light-cone sum rules considering different scenarios on quantum numbers of these states are investigated. The results ruled out the possible identification of the states with $J^P = \frac{1}{2}^-$ (for $\Omega_c(3000),\Omega_c(3050)$) and $J^P = \frac{3}{2}^-$ (for $ \Omega_c(3066),\Omega_c(3090)$). In order to make a final decision about the quantum numbers of these states, the contributions of all existent states should be taken into account simultaneously. This point needs further refined analysis.

\section{Conclusion}
In conclusion, we calculate the mass of newly observed excited $\Omega_c$ baryons with $J^P = \frac{1}{2}^-$ and $J^P = \frac{3}{2}^-$ at LHCb. Our results show that the sum rules predict their mass successfully once these new states are assigned as negative parity. For establishing the spin-parity assignment after determination of mass and residue it is necessary to calculate the decay widths and compare the result with the existing experimental data. Only after these comparisons, one can determine the spin-parity content of these newly observed states.

\textbf{Note added.---} While we were completing this study, the work~\cite{Wang:2017zjw} appeared in arXiv where new observed $\Omega_c$ particles are assigned as negative parity baryons and their masses are studied within QCD sum rules by using the different forms of interpolating currents than the ones we have used. Our results on mass are very close to the ones presented in~\cite{Wang:2017zjw}.

\bibliographystyle{ws-mpla}
\bibliography{T_AlievR2}

\begin{thebibliography}{10}

\bibitem{Aaij:2017nav}
LHCb Collaboration, R.~Aaij {\em et~al.}, {\em Phys. Rev. Lett.} {\bf 118},
  182001  (2017), \href{http://arxiv.org/abs/1703.04639}{{\ttfamily
  arXiv:1703.04639 [hep-ex]}}.

\bibitem{Yelton:2017qxg}
Belle Collaboration, J.~Yelton {\em et~al.}, {\em Phys. Rev.} {\bf D97},
  051102  (2018), \href{http://arxiv.org/abs/1711.07927}{{\ttfamily
  arXiv:1711.07927 [hep-ex]}}.

\bibitem{Yelton:2017uzv}
Belle Collaboration, J.~Yelton {\em et~al.}, {\em Phys. Rev.} {\bf D97},
  032001  (2018), \href{http://arxiv.org/abs/1712.01333}{{\ttfamily
  arXiv:1712.01333 [hep-ex]}}.

\bibitem{Wang:2017zjw}
Z.-G. Wang, {\em Eur. Phys. J.} {\bf C77},   325  (2017),
  \href{http://arxiv.org/abs/1704.01854}{{\ttfamily arXiv:1704.01854
  [hep-ph]}}.

\bibitem{Wang:2017xam}
Z.-G. Wang, X.-N. Wei and Z.-H. Yan, {\em Eur. Phys. J.} {\bf C77},   832
  (2017), \href{http://arxiv.org/abs/1706.09401}{{\ttfamily arXiv:1706.09401
  [hep-ph]}}.

\bibitem{Agaev:2017jyt}
S.~S. Agaev, K.~Azizi and H.~Sundu, {\em EPL} {\bf 118},   61001  (2017),
  \href{http://arxiv.org/abs/1703.07091}{{\ttfamily arXiv:1703.07091
  [hep-ph]}}.

\bibitem{PhysRevD.95.094018}
H.-Y. Cheng and C.-W. Chiang, {\em Phys. Rev. D} {\bf 95},   094018 (May 2017).

\bibitem{Kim:2017jpx}
H.-C. Kim, M.~V. Polyakov and M.~Praszałowicz, {\em Phys. Rev.} {\bf D96},
  014009  (2017), \href{http://arxiv.org/abs/1704.04082}{{\ttfamily
  arXiv:1704.04082 [hep-ph]}}, [Addendum: Phys. Rev.D96,no.3,039902(2017)].

\bibitem{Wang:2017vnc}
W.~Wang and R.-L. Zhu, {\em Phys. Rev.} {\bf D96},   014024  (2017),
  \href{http://arxiv.org/abs/1704.00179}{{\ttfamily arXiv:1704.00179
  [hep-ph]}}.

\bibitem{Karliner:2017kfm}
M.~Karliner and J.~L. Rosner, {\em Phys. Rev.} {\bf D95},   114012  (2017),
  \href{http://arxiv.org/abs/1703.07774}{{\ttfamily arXiv:1703.07774
  [hep-ph]}}.

\bibitem{Padmanath:2017lng}
M.~Padmanath and N.~Mathur, {\em Phys. Rev. Lett.} {\bf 119},   042001  (2017),
  \href{http://arxiv.org/abs/1704.00259}{{\ttfamily arXiv:1704.00259
  [hep-ph]}}.

\bibitem{Wang:2017hej}
K.-L. Wang, L.-Y. Xiao, X.-H. Zhong and Q.~Zhao, {\em Phys. Rev.} {\bf D95},
  116010  (2017), \href{http://arxiv.org/abs/1703.09130}{{\ttfamily
  arXiv:1703.09130 [hep-ph]}}.

\bibitem{PhysRevD.95.114024}
Z.~Zhao, D.-D. Ye and A.~Zhang, {\em Phys. Rev. D} {\bf 95},   114024 (Jun
  2017).

\bibitem{Chen:2017gnu}
B.~Chen and X.~Liu, {\em Phys. Rev.} {\bf D96},   094015  (2017),
  \href{http://arxiv.org/abs/1704.02583}{{\ttfamily arXiv:1704.02583
  [hep-ph]}}.

\bibitem{Huang:2017dwn}
H.~Huang, J.~Ping and F.~Wang, {\em Phys. Rev.} {\bf D97},   034027  (2018),
  \href{http://arxiv.org/abs/1704.01421}{{\ttfamily arXiv:1704.01421
  [hep-ph]}}.

\bibitem{Aliev:2018uby}
T.~M. Aliev, S.~Bilmis and M.~Savci, {\em Adv. High Energy Phys.} {\bf 2018},
  3637824  (2018), \href{http://arxiv.org/abs/1805.02964}{{\ttfamily
  arXiv:1805.02964 [hep-ph]}}.

\bibitem{PhysRevD.95.094008}
H.-X. Chen, Q.~Mao, W.~Chen, A.~Hosaka, X.~Liu and S.-L. Zhu, {\em Phys. Rev.
  D} {\bf 95},   094008 (May 2017).

\bibitem{Yang:2017rpg}
G.~Yang and J.~Ping  (2017), \href{http://arxiv.org/abs/1703.08845}{{\ttfamily
  arXiv:1703.08845 [hep-ph]}}.

\bibitem{Bagan:1991sc}
E.~Bagan, M.~Chabab, H.~G. Dosch and S.~Narison, {\em Phys. Lett.} {\bf B278},
  367  (1992).

\bibitem{Aliev:2002ra}
T.~M. Aliev, A.~Ozpineci and M.~Savci, {\em Phys. Rev.} {\bf D66},   016002
  (2002), \href{http://arxiv.org/abs/hep-ph/0204035}{{\ttfamily
  arXiv:hep-ph/0204035 [hep-ph]}}, [Erratum: Phys. Rev.D67,039901(2003)].

\bibitem{Wang:2008vg}
L.~Wang and F.~X. Lee, {\em Phys. Rev.} {\bf D78},   013003  (2008),
  \href{http://arxiv.org/abs/0804.1779}{{\ttfamily arXiv:0804.1779 [hep-ph]}}.

\bibitem{Aliev:2008sk}
T.~M. Aliev, K.~Azizi and A.~Ozpineci, {\em Nucl. Phys.} {\bf B808}, 137
  (2009), \href{http://arxiv.org/abs/0807.3481}{{\ttfamily arXiv:0807.3481
  [hep-ph]}}.

\bibitem{Aliev:2010yx}
T.~M. Aliev, K.~Azizi and M.~Savci, {\em Phys. Lett.} {\bf B696}, 220  (2011),
  \href{http://arxiv.org/abs/1009.3658}{{\ttfamily arXiv:1009.3658 [hep-ph]}}.

\end{thebibliography}
\vspace{5cm}

\clearpage
  
\appendix
\section{ Mass sum rules for the spin $1/2$ $(3/2)$ negative parity $\Omega_c$ ($\Omega_c^{*})$ baryon}
\label{sec:appendix}
In this appendix, we present the expressions of the invariant function $\Pi_1^B~(\Pi_1^{*B})$ and $\Pi_2^{B}~(\Pi_2^{*B})$ appearing in the mass sum rules for the $\Omega_c~(\Omega_c^{*})$ baryon. For brevity we did not present the terms proportional to the strange quark mass here, but in the numerical calculations we take these terms into account.

\begin{equation}
  \begin{split}
\Pi_1^{B} =&
-{M^{6} \over 256 \pi^4} \Big\{3 [5 + \beta (2 + 5 \beta)] m_c^4 ({\cal I}_3
- 2 m_c^2 {\cal I}_4 + m_c^4 {\cal I}_5)\Big\} \\
&- {M^2\over 3072 \pi^4}
\Big\{m_c^2 \Big([13 + \beta (10 + 13 \beta)] \GG  {\cal I}_2 -
   16 (1 + \beta + \beta^2) \GG m_c^2 {\cal I}_3  \\
&+ 576 (1 - \beta^2) m_c \pi^2
    \sp ({\cal I}_2 - m_c^2 {\cal I}_3)\Big)\Big\}  \\
&-{e^{-m_c^2/M^2}\over 73728 m_c M^2 \pi^4}
\Big[(1 + \beta)^2 \GG^2 m_c - 768 (1 - \beta)^2 m_0^2 m_c \pi^4 \sp^2  \\
&- 16 (1 - \beta^2) \GG \pi^2 \sp \left(7 m_0^2 - 12 m_c^2 e^{m_c^2/M^2}
{\cal I}_1\right)\Big]  \\
&- {e^{-m_c^2/M^2}\over 18432 M^{4} \pi^2}
(1 - \beta) m_0^2 m_c \Big[2 (1 + \beta) \sp\GG -                      
   384 (1 - \beta) m_c \pi^2 \sp^2\Big]  \\
&+ {e^{-m_c^2/M^2}\over 1728 M^{6}}  
(1 - \beta)^2 \GG m_c^2 \sp^2 +
{e^{-m_c^2/M^2}\over 1728 M^{8}}
(1 - \beta)^2 \GG m_0^2 m_c^2 \sp^2  \\
&- {e^{-m_c^2/M^2}\over 3456 M^{10}}
(1 - \beta)^2 \GG m_0^2 m_c^4 \sp^2 
-{e^{-m_c^2/M^2}\over 24} (1 - \beta)^2 \sp^2 \\
&+ {e^{-m_c^2/M^2}\over 384 m_c \pi^2}
(1 - \beta^2) \sp \Big[\GG\left(1 - 3 m_c^2 e^{m_c^2/M^2} {\cal I}_2\right) \\
&- 3 m_0^2 m_c^2 e^{m_c^2/M^2} \left(6 {\cal I}_1 -
13 m_c^2 {\cal I}_2\right)\Big] 
\end{split}
\end{equation}


\begin{equation}
\begin{split}
\Pi_2^{B} =&
- {M^{6} \over 256 \pi^4} 
3 (1 - \beta)^2 m_c^3 \left({\cal I}_2 - 2 m_c^2 {\cal I}_3 +
m_c^4 {\cal I}_4\right)   \\
&- {M^{4} \over 3072 \pi^4} 
(1 - \beta) m_c \Big\{3 (1 - \beta) \GG {\cal I}_2 - 
   4 m_c^2 \Big[(1 - \beta) \GG - 144 (1 + \beta) m_c \pi^2 \sp\Big]
{\cal I}_3\Big\}   \\
&+ {M^2 e^{-m_c^2/M^2}\over 1024 \pi^4}
(1 - \beta) \Big\{56 (1 + \beta) m_0^2 \pi^2 \sp -     
   2 (1 - \beta) \GG m_c e^{m_c^2/M^2} {\cal I}_1   \\
&+ m_c^2 e^{m_c^2/M^2} \Big[3 (1 - \beta) \GG m_c - 8 (1 + \beta) m_0^2
\pi^2 \sp\Big]
    {\cal I}_2\Big\}   \\
&- {e^{-m_c^2/M^2}\over 73728 M^2 \pi^4}
m_c \Big\{(1 - \beta)^2 \GG^2 + 1536 [3 + \beta (2 + 3 \beta)] m_0^2
\pi^4 \sp^2\Big\}   \\
&+ {e^{-m_c^2/M^2}\over 18432 M^{4} \pi^2}
m_c \Big\{22 (1 - \beta^2) \sp \GG m_0^2 m_c  \\
&- 32 [5 + \beta (2 + 5 \beta)] \left(\GG - 12 m_0^2 m_c^2\right)
    \pi^2 \sp^2\Big\}  \\
&- {e^{-m_c^2/M^2}\over 1728 M^{6}}
[5 + \beta (2 + 5 \beta)] \GG m_c (3 m_0^2 - m_c^2) \sp^2   \\
&+ {e^{-m_c^2/M^2}\over 576 M^{8}}
[5 + \beta (2 + 5 \beta)] \GG m_0^2 m_c^3 \sp^2  \\
&- {e^{-m_c^2/M^2}\over 3456 M^{10}}
[5 + \beta (2 + 5 \beta)] \GG m_0^2 m_c^5 \sp^2  \\
&+ {e^{-m_c^2/M^2}\over 36864 m_c \pi^4}
\Big\{(1 - \beta)^2 \GG^2 - 1536 [5 + \beta (2 + 5 \beta)] m_c^2 \pi^4 \sp^2 \\
&- 192 (1 - \beta^2) \GG m_c \pi^2 \sp\Big\} 
\end{split}
\end{equation}
%
%
%
%
\begin{equation}
  \begin{split}
\Pi_1^{*B} =&
{M^{6} \over 32 \pi^4}
m_c^4 \left({\cal I}_3 - 3 m_c^4 {\cal I}_5 + 2 m_c^6 {\cal I}_6\right)   \\
&- {M^2\over 1152 \pi^4}
m_c^2 \Big[192 m_c \pi^2 \sp \left({\cal I}_2 - m_c^2 {\cal I}_3\right) -
   \GG (4 {\cal I}_2 - 3 m_c^4 {\cal I}_4)\Big]   \\
&+ {e^{-m_c^2/M^2}\over 82944 m_c M^2 \pi^4}
\Big[96 \GG m_0^2 \pi^2 \sp - 4608 m_0^2 m_c \pi^4 \sp^2 +  
  \GG^2 \left(m_c + 2 m_c^3 e^{m_c^2/M^2} {\cal I}_2\right)\Big]   \\
&- {e^{-m_c^2/M^2}\over 1728 M^{4} \pi^2}
\Big[m_0^2 m_c \sp \left(\GG + 96 m_c \pi^2 \sp\right)\Big]   \\
&- {e^{-m_c^2/M^2}\over 648 M^{6}} \GG m_c^2 \sp^2
- {e^{-m_c^2/M^2}\over 6488 M^{8}} \GG m_0^2 m_c^2 \sp^2
+ {e^{-m_c^2/M^2}\over 1296 M^{10}} \GG m_0^2 m_c^4 \sp^2   \\
&+ {e^{-m_c^2/M^2}\over 432 m_c \pi^2}
\sp \Big[\GG + 48 m_c \pi^2 \sp - 3 m_c^2 \left(\GG - 6 m_0^2 m_c^2\right)
e^{m_c^2/M^2} {\cal I}_2\Big]~.
\end{split}
\end{equation}


\begin{equation}
  \begin{split}
\Pi_2^{*B} =&
{M^{6} \over 96 \pi^4} 
\left(2 m_c^3 {\cal I}_2 - 3 m_c^5 {\cal I}_3 + m_c^9 {\cal I}_5\right)   \\
&+ {M^{4} \over 2304 \pi^4}
m_c  \Big[3 \GG {\cal I}_2 - m_c^4 \left(5 \GG + 384 m_c \pi^2 \sp\right)
{\cal I}_4\Big]   \\
&- {M^2 e^{-m_c^2/M^2}\over 1152 \pi^4}
 \Big[\GG m_c^3 e^{m_c^2/M^2} (2 {\cal I}_2 - m_c^2 {\cal I}_3) -   
   48 m_0^2 \pi^2 \sp (1 - m_c^4 e^{m_c^2/M^2} {\cal I}_3)\Big]   \\
&+ {e^{-m_c^2/M^2}\over 82944 M^2 \pi^4} \GG^2 m_c
- {e^{-m_c^2/M^2}\over 864 M^{4} \pi^2}  
m_c \sp \Big[\GG m_0^2 m_c - 12 \left(\GG - 12 m_0^2 m_c^2\right) \pi^2
\sp\Big]   \\   \\
&+ {e^{-m_c^2/M^2}\over 216 M^{6}}
\GG (3 m_0^2 m_c - m_c^3) \sp^2 
- {e^{-m_c^2/M^2}\over 72 M^{8}}  
\GG m_0^2 m_c^3 \sp^2
+ {e^{-m_c^2/M^2}\over 432 M^{10}} 
\GG m_0^2 m_c^5 \sp^2   \\
&+ {e^{-m_c^2/M^2}\over 82944 m_c \pi^4}
\Big[384 \GG m_c \pi^2 \sp + 27648 m_c^2 \pi^4 \sp^2 - 
\GG^2 \left(1 + 3 m_c^2 e^{m_c^2/M^2} {\cal I}_2\right)\Big]   ~.
\end{split}
\end{equation}

The functions ${\cal I}_n~(n=1,\cdots, 6)$
are defined as:
\begin{equation}
\label{nolabel}
{\cal I}_n = \int_{m_c^2}^{s_0} ds\, {e^{-s/M^2} \over s^n}~. 
\end{equation}
where $s_0$ is the value of the continuum threshold.

\end{document}